\documentclass{aa}  

\usepackage{graphicx}
\usepackage{txfonts}
\usepackage{xcolor}
\usepackage{xspace}
\usepackage{hyperref}
\hypersetup{
     colorlinks   = true,
     citecolor    = blue
}

\def \inte {{\em INTEGRAL}}
\def \swift {{\em Swift}}
\def \chandra {{\em Chandra}}
\def \xmm {{\em XMM-Newton}}
\def \maxi{{\em MAXI J0911$-$655}}
\def\swiftj{{\em Swift J0911.9$-$6452}}
\def \swiftxrt{{\em Swift-XRT}}
\def \nustar{{\em NuSTAR}}
\def \maxigsc{{\em MAXI/GSC}}

\def \atca{{\em ATCA}}

\begin{document}

 \title{Discovery of a new accreting millisecond X-ray pulsar in the globular cluster NGC 2808}

   \author{A. Sanna,
          \inst{1}
          A. Papitto\inst{2},
          L. Burderi\inst{1},
          E. Bozzo\inst{3},
          A. Riggio\inst{1},
          T. Di Salvo\inst{4},
          C. Ferrigno\inst{3},
          N. Rea\inst{5,6},
          R. Iaria\inst{4}
          }

   \institute{Dipartimento di Fisica, Universit\`a degli Studi di Cagliari, SP Monserrato-Sestu km 0.7, 09042 Monserrato, Italy\\
   		\email{andrea.sanna@dsf.unica.it}
   	\and
                INAF, Osservatorio Astronomico di Roma, Via di Frascati 33, I-00044, Monteporzio Catone (Roma), Italy
         \and
                ISDC Data Centre for Astrophysics, Chemin d'Ecogia 16, CH-1290 Versoix, Switzerland
         \and
                Universit\`a degli Studi di Palermo, Dipartimento di Fisica e Chimica, via Archirafi 36, 90123 Palermo, Italy
         \and
               Anton Pannekoek Institute for Astronomy, University of Amsterdam, Postbus 94249, NL-1090 GE Amsterdam, The Netherlands
         \and
               Institute of Space Sciences (CSIC-IEEC), Campus UAB, Carrer Can Magrans s/n, E-08193 Barcelona, Spain       	   
             }

   \date{Received -; accepted -}

  \abstract
   {We report on the discovery of coherent pulsations at a period of 2.9 ms from the X-ray transient \maxi{} in the globular cluster NGC 2808. We observed X-ray pulsations at a frequency of $\sim339.97$~Hz in three different observations of the source performed with \xmm{} and \nustar{} during the source outburst. This newly discovered accreting millisecond pulsar is part of an ultra-compact binary system characterised by an orbital period of $44.3$ minutes and a projected semi-major axis of $\sim17.6$~lt-ms. Based on the mass function we estimate a minimum companion mass of 0.024~M$_{\odot}$, which assumes a neutron star mass of 1.4~M$_{\odot}$ and a maximum inclination angle of $75^{\circ}$ (derived from the lack of eclipses and dips in the light-curve of the source). We find that the companion star's Roche-Lobe could either be filled by a hot ($5\times 10^{6}$ K) pure helium white dwarf  with a 0.028~M$_{\odot}$ mass (implying $i\simeq58^{\circ}$) or an old (>5 Gyr) brown dwarf with metallicity abundances between solar/sub-solar and mass ranging in the interval 0.065$-$0.085 (16 < $i$ < 21). During the outburst the broad-band energy spectra are well described by a superposition of a weak black-body component (kT$\sim$ 0.5~keV) and a hard cutoff power-law with photon index $\Gamma \sim$ 1.7 and cut-off at a temperature kT$_e\sim$ 130~keV. Up to the latest \swiftxrt{} observation performed on 2016 July 19 the source has been observed in outburst for almost 150 days, which makes \maxi{} the second accreting millisecond X-ray pulsar with outburst duration longer than 100 days.}


\titlerunning{new AXMP \maxi{}}
\authorrunning{Sanna et al.}

   \maketitle

\section{Introduction}

After almost two decades from the discovery of the first accreting millisecond X-ray pulsar \citep[SAX J1808.4$-$3658;][]{Wijnands1998a}, the sample of accreting rapidly-rotating neutron stars (NS) harboured in low mass X-ray binary systems has increased in number up to 18 \citep[see][for an extensive review on the topic]{Burderi13,Patruno12b}. The extremely short spin periods shown by the accreting millisecond X-ray pulsars (hereafter AMXP) are the result of long-lasting mass transfer from low mass companion stars through an accretion disc onto a slow-rotating NS \citep[also know as ``recycling scenario'';][]{Alpar82}. At the end of the mass transfer phase, a millisecond pulsar shining from the radio to the gamma-ray band, and powered by the rotation of its magnetic field, is expected to turn on. The close link shared by radio millisecond pulsars and AMXPs has been observationally confirmed by the transitional binary systems IGR J18245$-$2452 \citep{Papitto2013b}, PSR J0023+0038 \citep{Archibald2009a, Stappers2014a,Patruno2014a, Archibald2015a} and XSS J12270 \citep{deMartino2010a, Bassa2014a, Roy2014a, Papitto2015a}.

From the observation of the 15 AMXPs from which a state transition has not been observed yet \citep[but see][]{de-Ona-Wilhelmi2016a}, it can be highlighted that AMXPs have spin frequencies uniformly distributed in the range $\sim$180-600~Hz. The orbital periods are always very short (P$_{orb} < 12$ hrs), with the exception of the intermittent Aql X$-$1 that shows a $\sim 18$ hrs orbital period \citep{Welsh2000a}. Five AMXPs of the sample show extremely short period (P$_{orb} < 1$ hrs) and can be considered \emph{ultra-compact} binary systems. Orbital constrains allowed estimates of the typical mass values of the companion stars of AMXPs to be made, showing the very small donors are usually preferred (consistent with donor masses below 0.2~M$_{\odot}$). Almost 80\% of the sources in the sample show persistent X-ray pulsation during the outburst phase. The three remaining sources have been observed pulsating only occasionally: Aql X$-$1 \citep{Casella08}, HETE J1900.1$-$2455 \citep{Kaaret06} and SAX J1748.9$-$2021 \citep{Gavriil07, Altamirano2008a, Patruno09a}. The latter, for which the X-ray pulsations turned on and off intermittently during the outbursts, has been recently observed pulsating during its latest outburst \citep{Sanna2016a}.

\maxi{} (also known as \swiftj{}) was firstly detected by the \maxigsc{} \citep{Mihara2011a} nova-alert system trigger the 19th of February 2016 \citep{Serino2016a} at a position compatible with the globular cluster NCG 2808. A few days later, on the 29th of February 2016, the {\em Burst Alert Telescope} \citep[BAT;][]{Barthelmy2005a} transient monitor onboard on \swift{} detected X-ray activity within a region compatible with the previous \maxigsc{} observation, confirming the detection of a new X-ray transient source. The source position was determined by \swift{} \citep{kennea2016a} and later on improved by \chandra{} \citep{Homan2016a}. Here we consider the most accurate source position, RA = $09^h12^m2.43^s$  and DEC = $-64^\circ52^m6.4^s$, determined by the latter with a 90\% confidence level uncertainty of 0.6''. Nearly-simultaneous radio and X-ray observations of the source performed by \atca{} on 2016 April 6 failed to detect a radio counterpart \citep{Tudor2016a}. Here we report on the discovery of coherent ms X-ray pulsation from \maxi{}, and we describe the detailed analysis of \xmm{} and \nustar{} observations from which we derived the orbital solution for the pulsar.

\section[]{Observations and data reduction}
\subsection{\xmm{}}
\label{sec:XMM}
\begin{figure}
\centering
\includegraphics[width=0.48\textwidth]{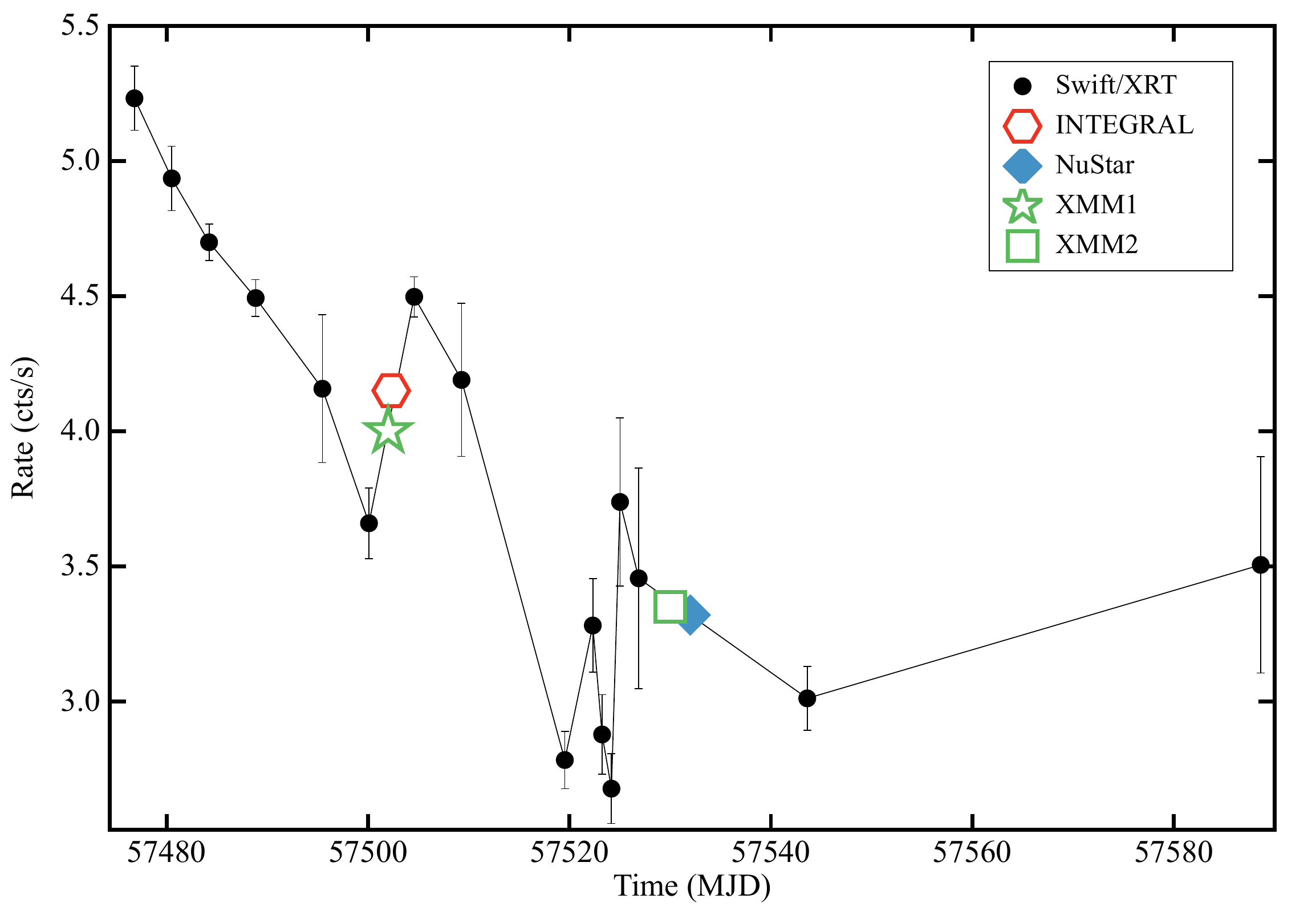}
\caption{Light-curve of the outburst of \maxi{} as observed by \swiftxrt{} (black points). Green symbols, blue diamond and red hexagon represent the observations collected by \xmm{}, \nustar{} and \inte{} respectively.}
\label{fig:lc}
\end{figure}

We analysed the pointed \xmm{} observations of \maxi{} performed on 2016 April 24 (Obs.ID. 0790181401, hereafter XMM1) and on 2016 May 22 (Obs.ID. 0790181501, hereafter XMM2). During both EPIC observations the pn camera was operated in timing mode (with an exposure time of $\sim$38 ks and $\sim$31 ks for XMM1 and XMM2, respectively), while the RGS instrument was observing in spectroscopy mode. Fig.~\ref{fig:lc} shows the light curve of observed outburst of the source monitored by \swiftxrt{} (black points). Each point represents the average count rate inferred from every pointed \swiftxrt{} observation following \citet{Evans2009a}\footnote{The mean count rate per observation have been extracted using the \swiftxrt{} on-line tool at  \url{http://www.swift.ac.uk}.}. The green star and the green rectangle represent observations XMM1 and XMM2, respectively. For this analysis we focused on the Epic-pn (PN) data, which have both statistics and time resolution (30~$\mu$s) required to investigate the time variability of the source. We performed the reduction of the PN data using the Science Analysis Software (SAS) v. 14.0.0 with the up-to-date calibration files, and adopting the standard reduction pipeline RDPHA \citep[see][for more details on the method]{Pintore14}. We filtered the PN data in the energy range 0.3$-$10.0 keV, selecting events with \textsc{pattern$\leq$4} with single and double pixel pattern only.
The observed PN mean count rate of the source extracted in the region RAWX range [31:45] during the observation XMM1 was $\sim40$~cts/s, which decreased to $\sim35$~cts/s during the observation XMM2. We estimated the background mean count rate in the RAWX range [2:6] to be $\sim0.15$~cts/s in the energy range 0.3$-$10~keV for both observations. We verified that the background region was not heavily contaminated by the source. During the observations no type-I burst episodes have been recorded.

We corrected the PN photon arrival times for the motion of the Earth-spacecraft system with respect to the Solar System barycentre by using the \textsc{barycen} tool (DE-405 solar system ephemeris). We used the best available source position obtained from the \textit{Chandra} identification of the source \citep[][]{Homan2016a}, and reported in Tab.~\ref{tab:solution}.

For each PN event file, we extracted energy spectra setting \textsc{`FLAG=0'} to retain only events optimally calibrated for spectral analysis. We generated response matrices and ancillary files using the \textsc{rmfgen} and \textsc{arfgen} tools, respectively. We grouped the PN energy channels in order to have not more than three bins per energy resolution elements, also we binned the energy spectra to have at least 25 counts per bin. We processed RGS data using the \textsc{rgsproc} pipeline, extracting first and second order spectra and response matrices. We binned the RGS spectra in order to have at least 25 counts per bin. 

\subsection{\inte{}}
\label{sec:integral}

\begin{figure}
  \includegraphics[width=0.49\textwidth]{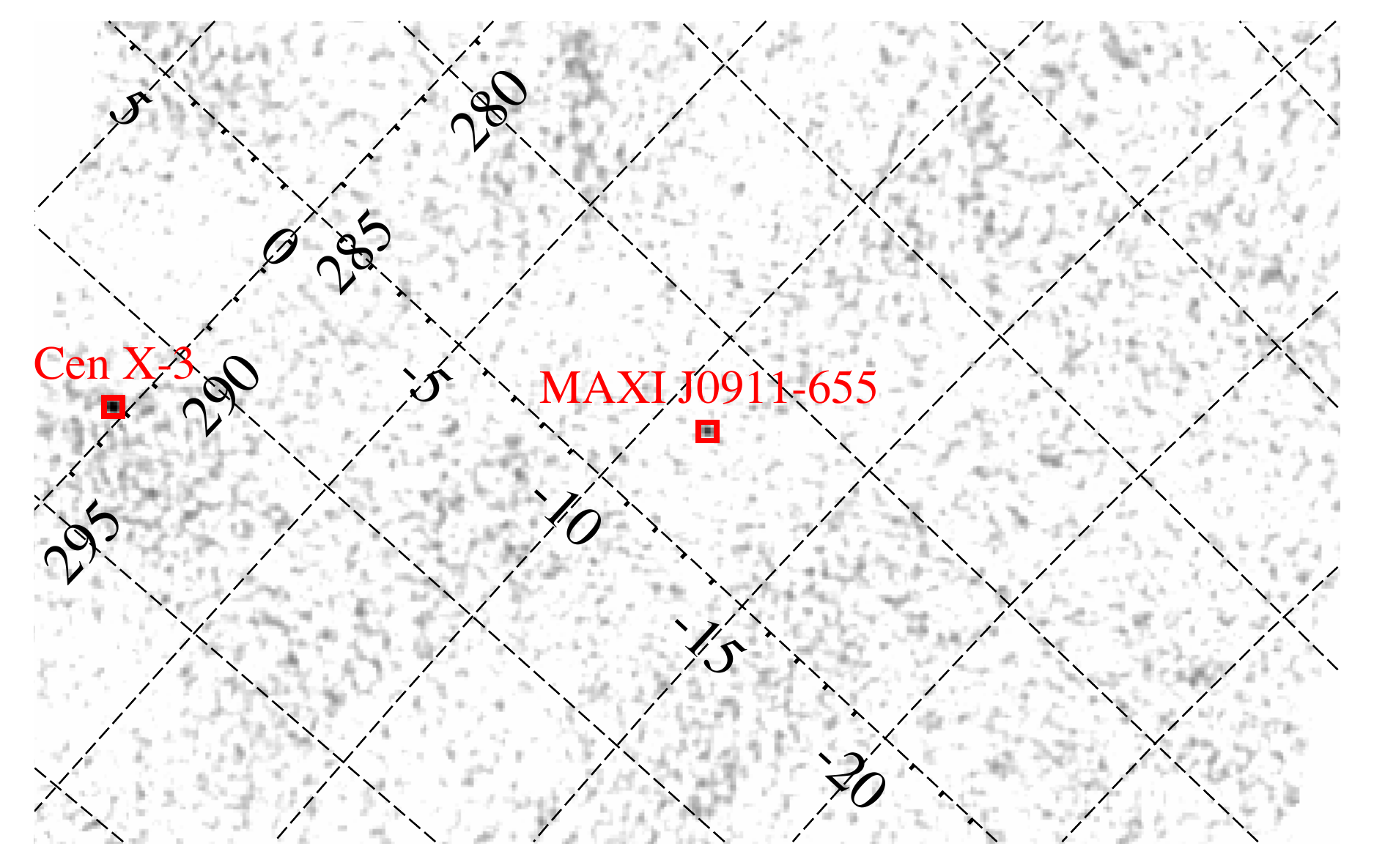}
  \caption{Mosaicked IBIS/ISGRI image around the position of \maxi{} obtained from the observations 
  performed in revolution 1671 (20-100~keV). The source is detected in this mosaic at a significance of 14~$\sigma$. Coordinates are expressed in degrees with respect to the galactic coordinate system.}   
  \label{fig:mosaic}
\end{figure}

\inte{} performed a dedicated target of opportunity observation of \maxi{} during the satellite revolution 1671, starting 
on 2016 April 23 at 23:43 and lasting until April 25 at 6:51 (red hexagon in Fig.~\ref{fig:lc}). The total exposure time collected in the 
direction of the source was of 71.5~ks. 
\noindent
\inte{} observations are divided into ``science windows'' (SCWs), 
i.e.,  pointings with typical durations of $\sim$2-3~ks. As the observation was carried out in the hexagonal mode, 
the source was always located close to the aim point of all on-board instruments.  
All data were analysed by using version 10.2 of the Off-line Scientific Analysis software   
(OSA) distributed by the ISDC \citep{courvoisier03}. We extracted the IBIS/ISGRI \citep{ubertini03,lebrun03} mosaic 
in the 20$-$100~keV energy band and the JEM-X \citep{lund03} mosaics in the 3$-$35~keV energy band by using all data 
available. The source was detected in the ISGRI mosaic at a significance of 14~$\sigma$ and a flux of 
5.9$\pm$0.4~mCrab\footnote{The conversion between the upper limit on the source count-rate and flux has been 
done by using the observations of the Crab in revolution 1597, as described in \citet{Bozzo2016a}}. This corresponds to 
roughly 1.1$\times$10$^{-10}$~erg~cm$^{-2}$~s$^{-1}$. Fig.~\ref{fig:mosaic} shows the mosaicked IBIS/ISGRI image around the position of \maxi{}, where the only other identifiable point source in the filed of view is Cen X-3. In the JEM-X mosaic the source was detected at 
9~$\sigma$ and a flux of 6.5$\pm$0.7~mCrab (i.e. roughly 2.0$\times$10$^{-10}$~erg~cm$^{-2}$~s$^{-1}$). 

As the source was relatively faint for \inte{}\ we extracted a single spectrum for IBIS/ISGRI and the two JEM-X units 
by using all the available exposure time (the effective exposure time was of 44~ks for ISGRI and 56~ks for both JEM-X1 and 
JEM-X2). 
Due to the relatively low statistics of the \inte{}\ data we did not attempt a detailed timing analysis. 
JEM-X light-curves with a time resolution of 2~s were extracted to search for type-I X-ray bursts, but no significant 
detection was found. 

\subsection{\nustar{}}
\maxi{} has been observed by \nustar{} (Obs.ID. 90201024002) on 2016 May 24 (blue diamond in Fig.~\ref{fig:lc}). We performed standard screening and filtering of the events by means of the \nustar{} data analysis software (\textsc{nustardas}) version 1.5.1, resulting in an exposure time of roughly 60~ks for each instrument. We selected source events extracting a circular region of radius 50'' centered in the source position. We used the same extracting region, but centered far from the source, to extract the background. Spectra and light curves from each instrument were extracted and response files were generated using the \textsc{nuproducts} pipeline. Furthermore, we used the \textsc{lcmath} task to generate background-subtracted light curves for the FMPA and FMPB, with an average count rate per instrument of $\sim4$ counts/s. No type-I burst episode have been recorded during the observation. 
We corrected the \nustar{} photon arrival times for the motion of the Earth-spacecraft system with respect to the Solar System barycentre by using the \textsc{barycorr} tool (DE-405 solar system ephemeris). We used the best available source position reported in Tab.~\ref{tab:solution}.

\section{Data analysis}

\subsection{Timing analysis}
\label{sec:ta2016}

We created a power density spectrum of the observation XMM1 by averaging together power spectra produced over 512 seconds of data. We detected a highly significant (13$\sigma$) broad double-picked signal with frequency ranging between 339.96 and 339.99 Hz. In order to obtain a first orbital motion solution for the NS we inspected every 512 s-long power density spectrum for significant features in the 0.04 Hz interval around the frequency 339.98 Hz. We modelled the detected spin frequency variation of the signal as the Doppler shift induced by the binary orbital motion:
\begin{equation}
\Delta_{\nu} = \Delta_{\nu_0} - \frac{2 \pi \, \nu_0 \, x}  {P_{orb}}  \cos\left(\frac{2 \pi \,(t - T_{NOD})} { P_{orb}}\right)
\end{equation}
where $\nu_0$ is the spin frequency, $x$ is the projected semi-major axis of the NS orbit in light seconds, $P_{orb}$ is the orbital period, and $T_{NOD}$ is the time of passage through the ascending node. 
We thus obtain a first estimate of the orbital ephemeris of the system, such as the orbital period $P_{orb}=2656(9)$ seconds, the projected semi-major axis $x = 0.0175(7)$ lt-s, $T_{NOD}=57502.185(1)$ (MJD), as well as the spin frequency $\nu_0=339.9751(5)$ Hz.

Starting from this first timing solution, we corrected the photon time of arrivals of the XMM1 datasets for the delays caused by the binary motion applying the orbital parameters through the recursive formula  
\begin{eqnarray}
\label{eq:barygen} 
t + z(t) = t_{arr}
\end{eqnarray}
where $t$ is photon emission time, $t_{arr}$ is the photon arrival time to the Solar System barycentre, $z(t)$ is the projection along the line of sight of the distance between the NS and the barycenter of the binary system in light seconds. For almost circular orbits (eccentricity $e \ll 1$) we can write
\begin{eqnarray} 
\label{eq:bary}
z(t)= x\,\sin\Big(\frac{2\pi}{P_{orb}} \,(t-T_{NOD})\Big)
\end{eqnarray}
The correct emission times (up to an overall constant $D/c$, where $D$ is the distance between the Solar System barycenter and the barycenter of the binary system) are calculated by solving iteratively the aforementioned Eq.~\ref{eq:barygen}, $t_{n+1} = t_{arr} - z(t_{n})$,
with $z(t)$ defined as in Eq.~\ref{eq:bary}, with the conditions $D/c = 0$, and $z(t_{n=0}) = 0$. We iterated until the difference between two consecutive steps (${\Delta t}_{n+1} = t_{n+1} - t_{n}$) was lower	 of the absolute timing accuracy of the instrument used for the observations. In our case we set ${\Delta t}_{n+1}=1 ~\mu$s.
\noindent
We folded 300 second intervals of the XMM1 dataset into 16 phase bins around the estimate of the pulsar spin frequency reported above. We fitted each folded profile with a sinusoid in order to determine the corresponding sinusoidal amplitude and the fractional part of the epoch-folded phase residual. We considered only folded profiles for which the ratio between the amplitude of the sinusoid and its 1~$\sigma$ uncertainty was larger than three. We tried to fit the the folded profiles including a second harmonic component, but this component was significantly detected in less than 5\% of the total number of intervals.

We modelled the temporal evolution of the pulse phase delays with the relation:
\begin{equation}
\label{eq:ph}
\Delta \phi(t)=\phi_0+\Delta \nu_0\,(t-T_0)+\frac{1}{2}\dot{\nu}\,(t-T_0)^2+R_{orb}(t)
\end{equation}
where $T_0$ represents the reference epoch for the timing solution, $\Delta \nu_0=(\nu_0-\bar{\nu})$ is the difference between the frequency at the reference epoch and the spin frequency used to epoch-fold the data, $\dot{\nu}$ is the spin frequency derivative, and $R_{orb}$ is the Roemer delay caused by the differential corrections to the set of orbital parameters used to correct the photon time of arrivals \citep[see e.g.][]{Deeter81}. If a new set of orbital parameters was found, photon time of arrivals were corrected using Eq.~\ref{eq:bary} and pulse phase delays were created and modelled with Eq.~\ref{eq:ph}. We iterated this process until no significant differential corrections were found for the parameters of the model. Obtained best-fit parameters are shown in the first column of Tab.~\ref{tab:solution}.
\noindent
Using the aforementioned NS ephemeris, we corrected the photon time of arrivals of the second \xmm{} observation (XMM2) by using Eq.~\ref{eq:bary}. We calculated pulse phase delays over pulse profiles created folding 300 second data intervals fitted with a sinusoidal profile. No second harmonic component was required to model the pulse profiles. Using Eq.~\ref{eq:ph} we calculated the spin frequency and the best-fit orbital parameters for XMM2 (see second column of Tab.~\ref{tab:solution}).

\begin{figure}
\centering
\includegraphics[width=0.45\textwidth]{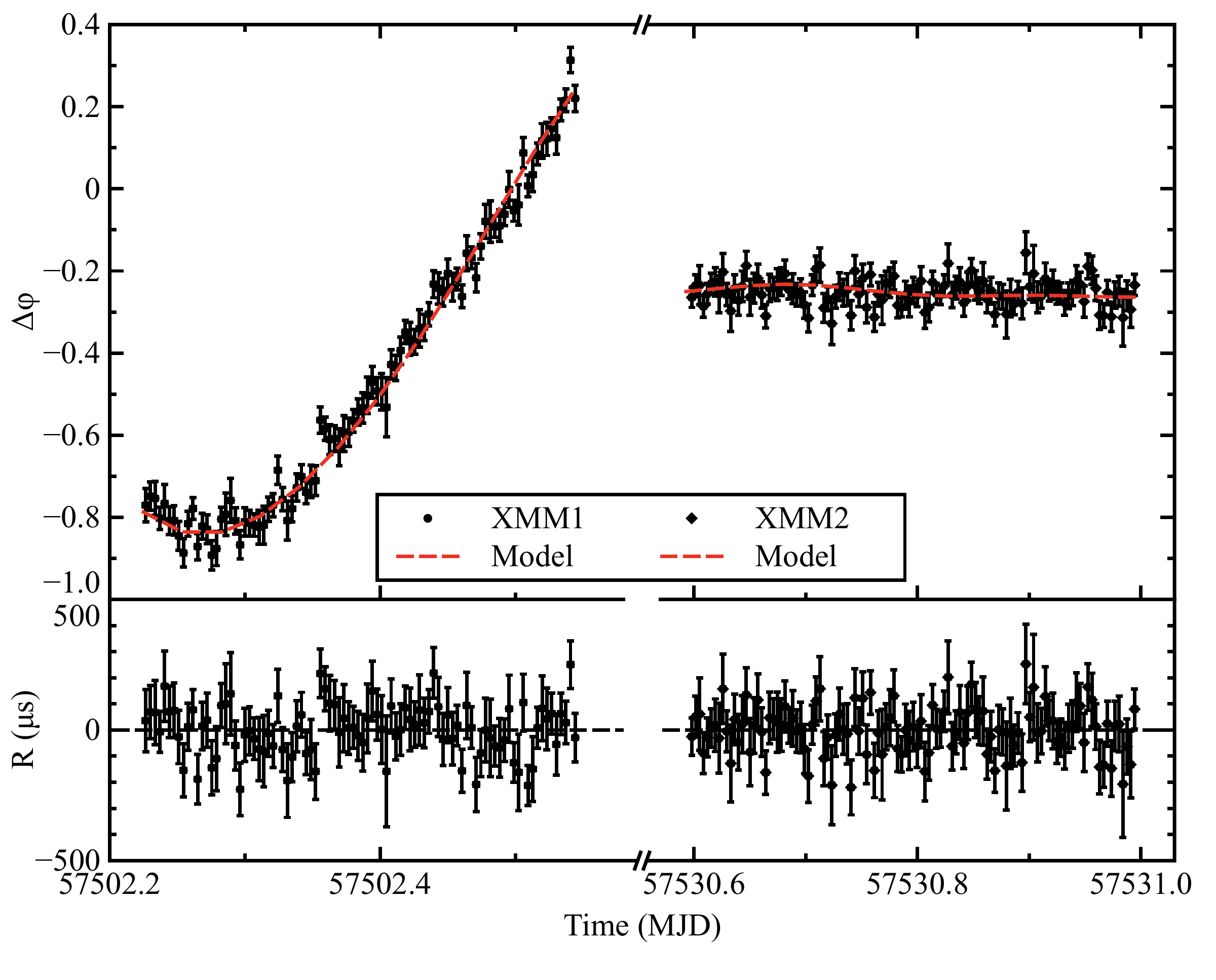}
\caption{\textit{Top panel -} Pulse phase delays as a function of time computed by epoch-folding 300 second data intervals of the \textit{XMM-Newton} observations, together with the best-fit model (red dotted line, see text). \textit{Bottom panel -} Residuals in $\mu s$ with respect to the best-fitting orbital solution.}
\label{fig:phase_fit}
\end{figure}

Uncertainties on the NS spin frequencies combined with the almost 30-days time gap between the two \xmm{} observations did not allow us to phase-connected the pulsations obtained from the two datasets. Still we were able to model together the phase delays induced by the orbital motion in the two observations. After correcting the photon time of arrivals of the XMM1 and XMM2 datasets using the orbital parameters determined from the timing analysis of XMM1, we calculated pulse phase delays following the process described previously.
We investigated differential correction to the adopted orbital parameters by fitting the pulse phase delays time evolution with the following models:
\begin{eqnarray}
\label{eq:ph2}
\begin{cases}
\Delta \phi_{XMM1}(t)=\sum\limits_{n=0}^{3} \frac{C_n}{n!}(t-T_0)^n+R_{orb}(t)\\
\Delta \phi_{XMM2}(t)=\sum\limits_{n=0}^{1} \frac{D_n}{n!}(t-T_0)^n+R_{orb}(t)
\end{cases}
\end{eqnarray}
where the first element of both equations represents a polynomial function used to model independently phase variations in each dataset. Additionally the model includes a common residual orbital modulation component $R_{orb}(t)$, which implies that the orbital parameters will be linked during the fit of the two dataset. This method has the advantage (with respect to model each observation separately) to improve the accuracy of the orbital parameters (e.g. the orbital period). Best-fit parameters are reported in Tab.~\ref{tab:solution}, while in Fig.~\ref{fig:phase_fit} we showed the pulse phase delays of the two observations with the best-fitting models (top panel), and the residuals with respect to the models.

In order to take into account the effect of the uncertainties on the source coordinates on the pulse phase delays, we used the expression of the residuals induced by the motion of the Earth for small variations of the source position $\delta_{\lambda}$ and $\delta_{\beta}$ expressed in ecliptic coordinates $\lambda$ and $\beta$ \citep[see, e.g.][]{Lyne90} to estimate the systematic uncertainties induced on the linear and quadratic terms of the pulse phase delays, which correspond to a systematic uncertainty in the spin frequency correction $\Delta \nu_0$, and the spin frequency derivative $\dot{\nu}$, respectively. The former corrections can be expressed as $\sigma_{\nu_{pos}}\leq \nu_0\,y\,\sigma_{\gamma}(1+\sin^2\beta)^{1/2}2\pi/P_{\oplus}$, and $\sigma_{\dot{\nu}_{pos}}\leq \nu_0\,y\,\sigma_{\gamma}(1+\sin^2\beta)^{1/2}(2\pi/P_{\oplus})^2$, where $y=r_E/c$ is the semi-major axis of the orbit of the Earth in light-seconds, $P_{\oplus}$ is the Earth orbital period, $\sigma_{\gamma}$ is the positional error circle and $\beta$ corresponds to  $\sim-70.7$ degrees. Considering the 90\% positional uncertainty of $0.6''$ reported by \citet{Homan2016a}, we estimated $\sigma_{\nu_{pos}} \leq 9\times 10^{-8}$~Hz, and $\sigma_{\dot{\nu}_{pos}} \leq 2\times 10^{-14}$~Hz/s, respectively. We added in quadrature these systematic uncertainties to the statistical errors of $\nu_0$, and $\dot{\nu}$ reported in Table.~\ref{tab:solution}.

Finally, using the binary parameters inferred from the timing analysis of the combined \xmm{} observations, we corrected the photon arrival times of the \nustar{} observation. We look for pulsations performing epoch-folding search of the whole observation using 16 phase bins. Starting from the spin frequency measured during the observation XMM2 (closest in time) we explored the frequency space with steps of $10^{-7}$~Hz for a total of 5001 steps. We detected a significant X-ray pulsation at a mean frequency of $\nu_{nus}=339.974925(3)$~Hz\footnote{The error on the spin frequency value has been estimated by means of Monte Carlo simulations.}, which significantly differs from the starting seed spin frequency. To investigate the timing properties of the source during the \nustar{} observation we calculated the pulse phase delays epoch-folding time intervals of approximately 6000 seconds (shortest time interval required to detect a significant pulse profile). We modelled each epoch-folded pulse profile with a sinusoid in order to determine the corresponding sinusoidal amplitude and the fractional part of phase residual, no second harmonic was required to model the profile. The length of the pulse profiles compared with the orbital binary period clearly excludes the possibility to investigate differential correction to the set of orbital parameters used to correct the photon arrival times. We therefore modelled the pulse phase delays with a polynomial function of the second order to investigate the NS spin frequency time evolution. Best-fit parameters are shown in Tab.\ref{tab:solution}.

\begin{table*}

\begin{tabular}{l | c  c  c  c}
Parameters             & XMM1 & XMM2 & XMM1+XMM2 & \nustar{} \\
\hline
\hline
R.A. (J2000) &  \multicolumn{4}{c}{$09^h12^m2.43^s$}\\
DEC (J2000) & \multicolumn{4}{c}{$-64^\circ52^m6.4^s$}\\
Orbital period $P_{orb}$ (s) &2659.71(14) &2659.88(7) & 2659.93312(47) & $^a$\\
Projected semi-major axis $x$ (lt-s) &0.01759(2) &0.017598(11) & 0.017595(9) & $^a$ \\
Ascending node passage $T_{NOD}$ (MJD) & 57502.185183(5) &57530.600905(13) &57502.185176(10)& $^a$\\
Eccentricity (e) &$ < 1.4 \times 10^{-2} $&$<4 \times 10^{-3}$ &$<5 \times 10^{-3}$ & $^a$\\
Spin frequency $\nu_0$ (Hz) &339.975071(3)&339.9750123(3) & $-$ & 339.974937(3)\\
Spin frequency derivative $\dot{\nu}$ (Hz/s) &-2.76(16)$\times 10^{-9}$& $-$& $-$ & -1.6(4)$\times 10^{-10}$\\
Epoch of $\nu_0$ and $\dot{\nu}$, $T_0$ (MJD) & 57502.2 &57530.6 & 57502.2 & 57532.0\\
\hline
$\chi^2_\nu$/d.o.f. & 141.4/86 &85.9/108 & 178.1/196  & 22.4/9\\
\end{tabular}
\caption{Orbital parameters and spin frequency of \maxi{} obtained from the analysis of the \xmm{} observations of the source. Errors are at 1$\sigma$ confidence level. The reported X-ray position of the source has a pointing uncertainty of 0.6$''$ \citep[see e.g.][]{Homan2016a}.$^a$This parameters has been fixed to the value obtained from the combined analysis of the \xmm{} observations.}
\label{tab:solution}
\end{table*}

\subsection{Spectral analysis}

The goal of the spectral analysis presented here is to characterise the X-ray emission of the newly discovered AMXP \maxi{} in a broad-band energy range. In order to do that we analysed the available X-ray observations of the source performed by different instruments, and we combined together observations close in time for which no spectral changes have been observed. We created two broad-band energy spectra, the first (hereafter Obs.1) includes the data collected by both the RGS and the PN instruments during the XMM1 observation, and the data collected by \inte{} that globally covered the time interval between 2016 April 24 and April 25. The second spectrum (hereafter Obs.2) includes the data collected by both the RGS and the PN instruments during the XMM2 observation, and the data from the \nustar{} observation performed almost two days after XMM2.

\begin{figure}
	\begin{center}$
		\begin{array}{c}
  			\includegraphics[width=0.45\textwidth]{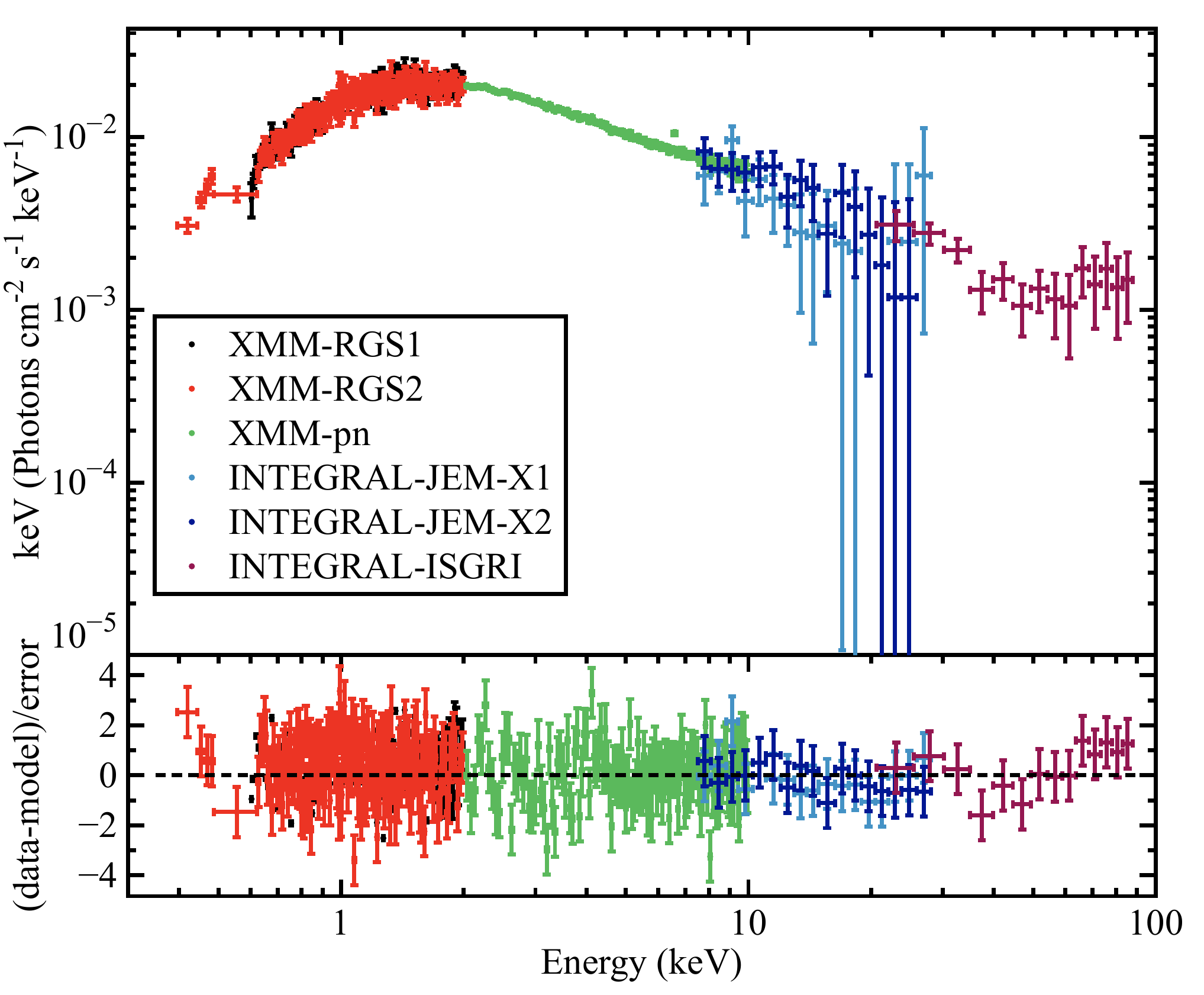}\\
 			 \includegraphics[width=0.45\textwidth]{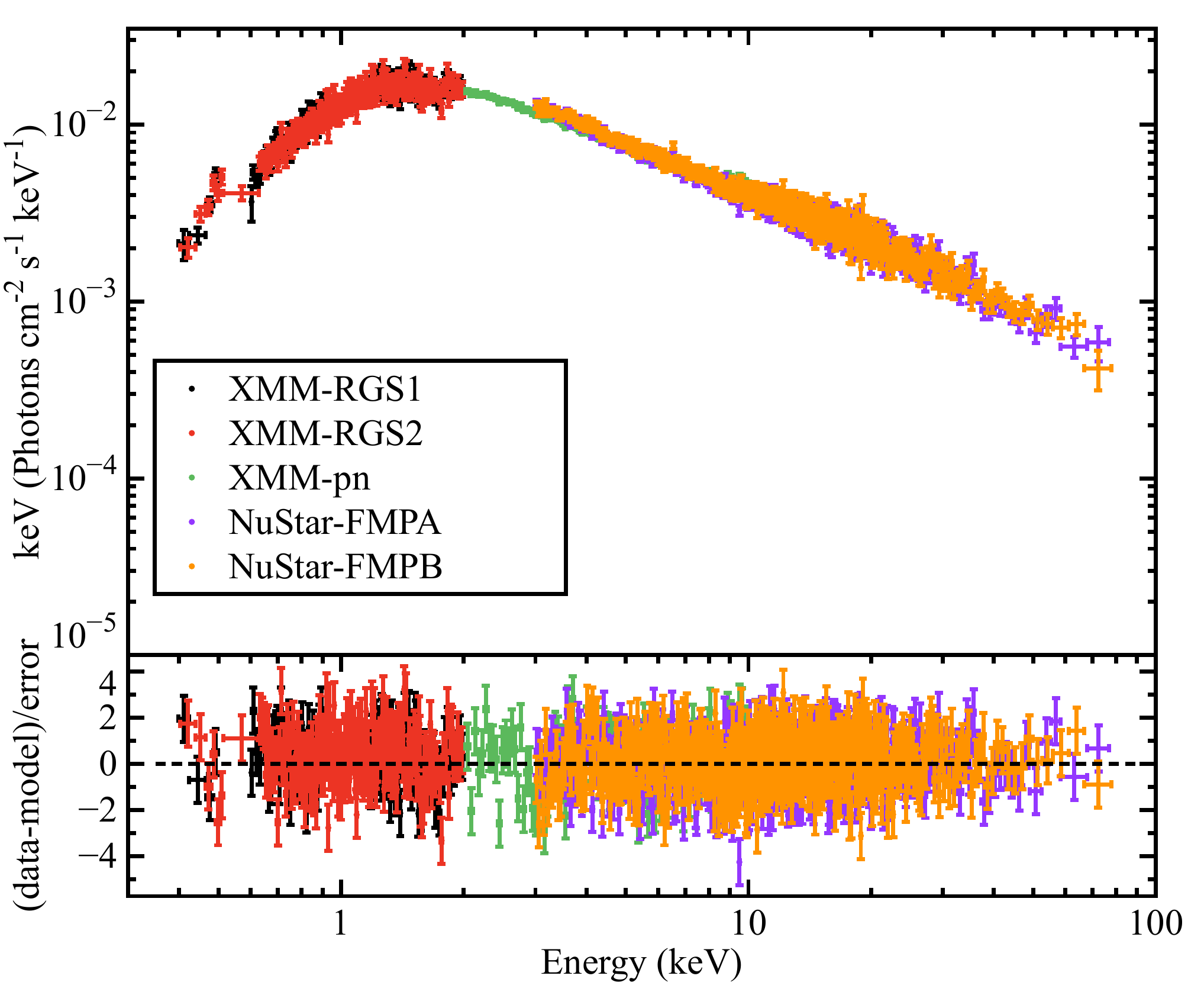}\\
  		\end{array}$
  	\end{center}
  	\caption{\textit{Top panel -} Broad-band (0.4$-$100~keV) energy spectrum of Obs.1 including \xmm{} RGS1 (black), RGS2 (red) and PN (green) and \inte{} JEM-X (blue and cyan) and IBIS/ISGRI. \textit{Bottom panel -} Broad-band (0.4$-$80~keV) energy spectrum of Obs.2 including \xmm{} RGS1 (black), RGS2 (red) and PN (green) and \nustar{} FMPA/B (orange and violet). For both energy spectra the residuals with respect to the best fit models are given in units of $\sigma$.}   
  	\label{fig:spectra}
\end{figure}

\noindent
Both broad-band spectra, reported in the top panel (0.4-100 keV Obs.1) and bottom panel (0.4-80 keV Obs.2) of Fig.~\ref{fig:spectra}, appeared to be quite hard. They are well fitted ($\widetilde{\chi}^2_\nu$/d.o.f. 1.04/1462 and 1.05/2700, for Obs.1 and Obs.2, respectively) by an absorbed power-law with high energy cutoffs, that resulted statistically more significant that a simple power-law component ($\Delta \chi^2 = 26$ and $\Delta \chi^2 = 360$ for the addition of one parameter, for Obs.1 and Obs.2, respectively). The component that takes into account interstellar absorption ({\sc Tbabs}) uses abundances and photoelectric cross-section table of \citet{Anders89} and \citet{Balucinska-Church92}, respectively. We find a neutral column density ($N_H$) of $(0.25\pm0.01) \times 10^{22}$ cm$^{-2}$ and $(0.29\pm0.01) \times 10^{22}$ cm$^{-2}$, a photon index $1.59\pm0.06$ and $1.75\pm0.02$, and a cutoff temperature of $88^{+120}_{-23}$ keV and $160_{-23}^{+32}$ keV, for Obs.1 and Obs.2, respectively. Furthermore, we find evidence for a weak soft thermal component that we fitted with a blackbody of temperature $0.61\pm0.1$ keV in Obs.1, that decreases to $0.53\pm0.01$ keV for Obs.2. Finally, both spectra show the presence of a weak (marginally significant) and relatively narrow emission line in the range 6.5-6.6 keV, that we model with a gaussian profile with $\sigma$ ranging between 0.02 and 0.2 keV and equivalent width between 11 and 16 eV, which we identified with K$\alpha$ emission from Fe XXV.

We estimated the broad-band 0.4$-$100 keV absorbed (unabsorbed) flux of Obs.1 and Obs.2 to be $\sim4.3\times 10^{-10}$ erg cm$^{-2}$ s$^{-1}$ ($\sim4.7\times 10^{-10}$ erg cm$^{-2}$ s$^{-1}$) and $\sim3\times 10^{-10}$ erg cm$^{-2}$ s$^{-1}$ ($\sim3.4\times 10^{-10}$ erg cm$^{-2}$ s$^{-1}$), respectively. Considering a source distance of $\sim$9.5 kpc \citep[distance of the host globular cluster NGC 2808;][]{Watkins2015a} we estimate an unabsorbed luminosity of $\sim5.1\times 10^{36}$ erg cm$^{-2}$, and $\sim3.7\times 10^{36}$ erg cm$^{-2}$, for Obs.1 and Obs.2, respectively, corresponding to 3\% and 2\% of the Eddington luminosity for accretion onto a standard 1.4 M$_\odot$ NS.

\section{Discussion}

We have discovered a new AMXP, \maxi{}, which is located in the globular cluster NGC 2808. X-ray pulsations at 339.97~Hz were detected in the two \xmm{} and in the \nustar{} observations performed during the outburst of the new X-ray source, with an average pulse fraction of 7\%. We modelled the NS spin frequency drift shown by the source as Doppler shift induced by the binary orbital motion, finding an accurate orbital solution of the almost 45 minutes binary system. We found an accurate set of orbital parameters by fitting the pulse phase delays of the two \xmm{} observations simultaneously, modelling the delays introduced by the orbital motion with a common set of parameters (orbitally phase-connected). On the other hand, the evolution with time of the spin NS spin frequency was kept independent between the two observations. The orbital period (44.3~minutes) and the binary separation (17.6~lt-ms) clearly point out the ultra-compact nature of  the binary system, which recall very similarly the orbital properties of other AMXPs such as XTE J1751$-$305 \citep{Markwardt02, Papitto08, Riggio11b}, XTE J0299$-$314 \citep{Galloway02}, XTE J1807$-$294 \citep{Kirsch04, Riggio08, Chou2008a, Patruno2010a}, SWIFT J1756.9$-$2508 \citep{Krimm07, Linares08, Patruno10b} and NGC6440 X$-2$ \citep{Altamirano2010a, Bult2015a}. 

\noindent
As reported in Tab.~\ref{tab:solution}, both XMM1 and \nustar{} observations required a very large spin-down NS frequency derivative in order to adequately model the evolution of the pulse phase delays. We note that both measurements are artefacts caused by instrumental issues of the satellites during the observations, more specifically the XMM1 observations showed anomalies on the Spacecraft Time Correlation file which where only partially corrected by the \xmm{} science team (\xmm{} team private communication), while the \nustar{} observation suffered from a time drift of the internal clock of the instrument \citep{Madsen15}. The presence of spurious spin frequency derivatives significantly changes the spin frequency measurements explaining the large discrepancy between the spin frequency detected during XMM2 and the other two observations. 

The latest \swift{} observation of \maxi{} performed on 2016 July 19, showed X-ray activity comparable with that observed during the \xmm{} and \nustar{} observations where the source has been observed pulsating. Assuming that the outburst started with the first detection of the source (2016 February 19) we can estimate an outburst durations of at least 150~days, which results quite long compared with the other AMXPs, although not unique. With the exception of the quasi-persistent AMXP HETE J1900.1$-$2455 (observed in outburst for more than 8 years), a long-lasting outburst has been observed for the ultra-compact AMXP XTE J1807$-$294, detected in outburst for almost 120 days during its 2003 outburst \citep{Riggio07}. We note however, that the lack of high timing resolution observations at the extremes of the observed outburst do not allow us to verify the persistency of the X-ray pulsation during the long outburst.

\begin{figure}
  \includegraphics[width=0.49\textwidth]{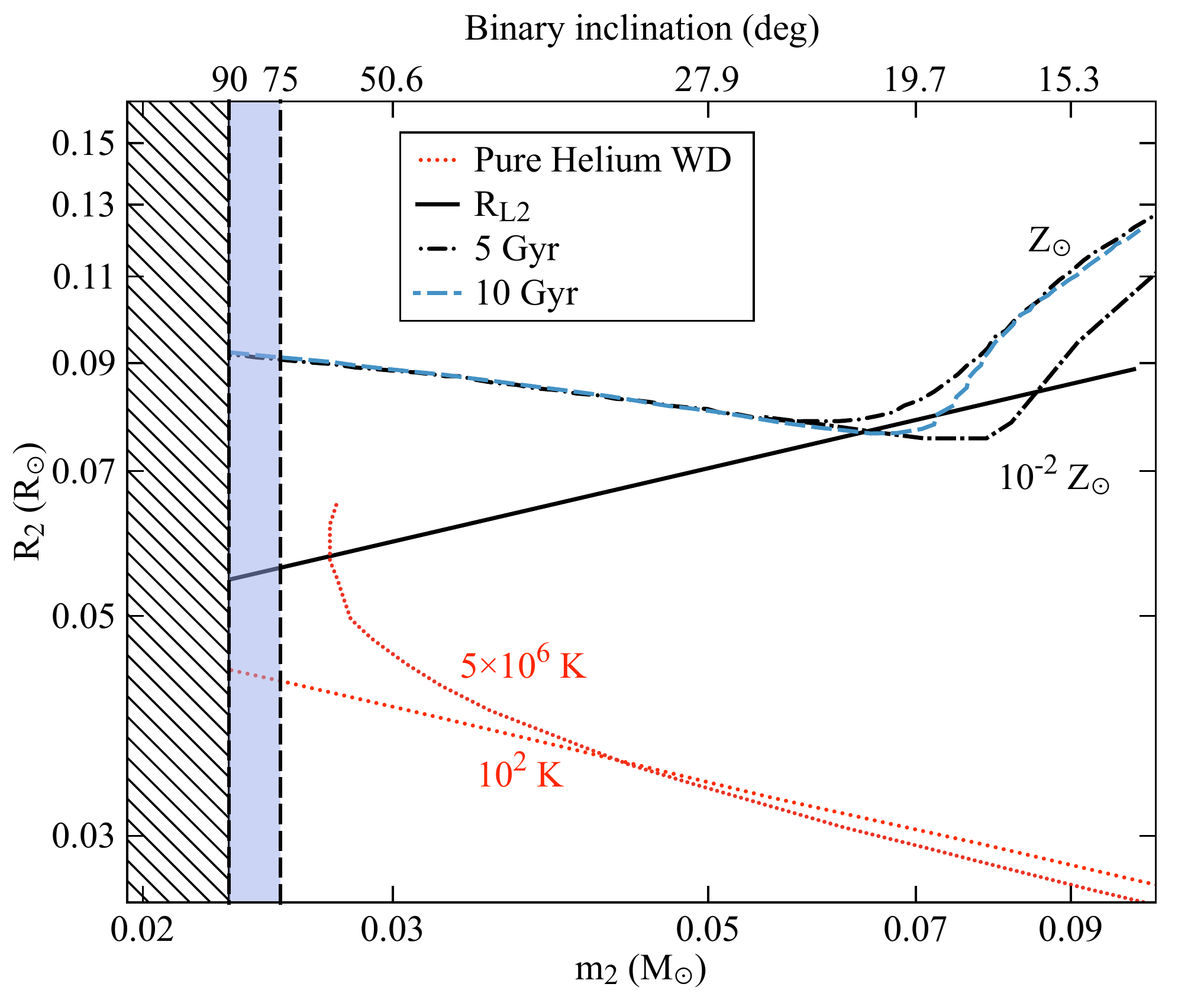}
  \caption{Radius-mass plane showing the size constraints on the companion star Roche-Lobe of \maxi{} (black solid line). The other curves represent theoretical mass-radius relations for hot ($5\times 10^6$ K) and cold ($10^2$ K) pure He white dwarfs (red dotted lines), low-mass main sequence/brown dwarfs of age 5 Gyr for solar and sub-solar metallicity abundances (black dot-dashed lines) and 10 Gyr (cyan dashed line). Top x-axes represents the corresponding binary inclination angle in degrees assuming a NS of 1.4 M$\odot$. The hatched region represents the constraints on the companion mass from the binary mass function, while the blue shaded area represents the mass constraints for inclination angles between 75 and 90 degrees.}     
  \label{fig:mass}
\end{figure}

The value of NS mass function $f(m_2, m_1, i)\sim6.2\times 10^{-6}$~M$_{\odot}$ allows us to set constrains on the mass of the companion star. The lack of eclipses, as well as dips, in the light curve sets an upper limit $i \lesssim 75^{\circ}$ on the inclination angle \citep{Frank02}, which provides a lower limit to the companion star of $m_2 \gtrsim 0.024$~M$_{\odot}$ (assuming a 1.4~M$_{\odot}$ NS), which increases up to $m_2 \gtrsim 0.03$~M$_{\odot}$ if we consider a 2~M$_{\odot}$ NS. Combining the condition for mass transfer via Roche-Lobe overflow ($R_2\approx R_{L2}$) with the mass function we can describe the companion mass radius as $R_2\simeq 0.2\,m_2^{1/3}\,P_{orb,1h}^{2/3}$ R$_{\odot}$, where P$_{orb,h}$ is the binary orbital period express in hours. In Fig.~\ref{fig:mass} we compare the companion mass-radius relation (black solid line) with detailed numerical simulated mass-radius relations for low-mass hydrogen main sequence stars and brown dwarfs of different age \citep[black dot-dashed line 5~Gyr and cyan dashed line 10~Gyr;][]{Chabrier2009a} and different metallicity content \citep[5~Gyr black dot-dashed line with solar metallicity Z$_{\odot}$ and sub-solar (0.01~Z$_{\odot}$) abundances;][]{Chabrier2000a}. From the intersections between mass-radius companion curve and the stellar models we can infer that the donor star is compatible with an old brown dwarf ($\geq$ 5~Gyr), with mass in the interval 0.065$-$0.085~M$_{\odot}$\footnote{Mass values within the intersections, for which the Roche-Lobe radius is larger than the estimated stellar radius, are still acceptable if we considered the possibility of the companion being bloated with respect to its thermal equilibrium radius because of irradiation from the compact object.} and metallicity abundances between solar and sub-solar. We note that age and metallicity content are compatible with estimates from deep IR studies of stellar populations of host cluster NGC 2808 \citep{Massari2016a}. The latter companion mass range limits the binary inclination angle between 16 and 21 degrees. In Fig.~\ref{fig:mass} we report also theoretical mass-radius relations for cold ($10^2$~K) and hot ($5\times 10^6$~K) pure helium white dwarfs \citep{Deloye2003a}. The latter object is compatible with the mass-radius companion curve for a mass value of $\sim 0.028$~M$_{\odot}$, which implies an inclination angle of $\sim 58$ degrees. Studies on evolutionary paths should be investigated in order to understand which of the two proposed scenarios is more favourable.  
 
Finally, the broad-band energy spectra of \maxi{} analysed here (Obs.1 and Obs.2) are well described by the superposition of a weak soft black-body like component ($kT\sim 0.5$~keV) and a hard high-energy cutoff power-law ($\Gamma \sim 1.7$ and kT$_e \sim$ 130~keV), in agreement with spectral properties other AMXPs observed in a hard state \citep{Falanga05a, Falanga05b, Gierlinski2005a, Patruno09b, Papitto09, Papitto2013a}. Moreover, the source shows a marginal evidence of a weak and narrow reflection component in the energy range 6.5$-$6.6~keV which we identify as the K$\alpha$ emission line from helium-like iron. The estimated broad-band (0.4$-$100~keV) unabsorbed flux of the source varies from $\sim 4.7\times 10^{-10}$~erg~s$^{-1}$ cm$^{-2}$ (Obs.1) to $\sim 3.4\times 10^{-10}$~erg~s$^{-1}$ cm$^{-2}$ (Obs.2) after almost a month of outburst decay (see Fig.~\ref{fig:lc}), which implies a bolometric luminosity of $\sim 5.1\times 10^{36}$~erg~s$^{-1}$ and $\sim 3.7\times 10^{36}$~erg~s$^{-1}$, respectively \citep[assuming the source located at 9.5~kpc;][]{Watkins2015a}. Considering $\tau=150$~day a good proxy for the outburst duration and assuming an outburst mean luminosity of the order of $\bar{L} = 4.5\times 10^{36}$ erg s$^{-1}$ (mean value between Obs.1 and Obs.2), we can estimate the amount of matter accreted into the NS during the outburst as:
\begin{equation}
\Delta M_{acc}\simeq 0.3 \times 10^{-10} \bar{L}_{36}\,\tau_{150}\,R_{1,10}\,m_{1,1.4}^{-1} \,~M_{\odot}
\end{equation}
where $\bar{L}_{36}$ is the mean bolometric luminosity in units of $10^{36}$ erg s$^{-1}$, $R_{1,10}$ and m$_{1,1.4}$ are the NS radius in units of 10~km and the NS mass in units of 1.4~M$_{\odot}$, respectively. Considering that in binary systems with a very low mass companions the mass transfer via the Roche-Lobe overflow is driven by angular momentum loss from GR \citep[see e.g.][]{Verbunt93}, we can write the long-term mass transfer rate as:
\begin{equation}
\dot{M}_{GR}\simeq 6.3\times 10^{-11}\, m_{1,1.4}^{2/3}\, m_{2,0.05}^{2}\, P_{orb,h}^{-8/3}\,M_{\odot}/yr
\label{eq:mass_acc}
\end{equation}
where m$_{2,0.05}$ represents the companion star mass in units of 0.05 M$_{\odot}$ and for which we assume that the companion star is well described by the relation $R_2\propto M_2^{-1/3}$ suited for degenerate or fully convective stars. Substituting the two sets of mass values for the companion discussed in the previous paragraph, we estimate $\dot{M}_{GR}\simeq 4.6\times 10^{-11}$ M$_{\odot}$/yr (for a hot pure helium white dwarf companion star) and $\dot{M}_{GR}\simeq 3.8\times 10^{-10}$ M$_{\odot}$/yr (assuming an old brown dwarf companion star). If all the matter transferred from the donor is accreted into the NS (conservative mass transfer), we can combine the two estimates of the mass accretion rate with Eq.~\ref{eq:mass_acc} to infer the amount of time required to transfer the matter accreted on the compact object and responsible for the observed outburst of \maxi{}. In analogy with the transient nature of the other known AMXPs, the inferred time interval could also be interpreted as a rough estimate of the outburst recurrence time. We obtain a time period of $\sim3$ years for a helium white dwarf companion star, and $\sim0.3$ years in the case of a brown dwarf companion star. If this is correct, even in the less favourable scenario, the source should show and it should have shown outburst episodes on the timescales of several years. The lack of observed X-ray outbursts from \maxi{} in the past 10 years, regardless the presence of high sensitivity all sky monitor systems (that indeed discovered the source as soon as it went in outburst) such as the \swift{} BAT, reveals inconsistencies in the scenario pictured before. An intriguing, although highly speculative at this stage, way out to reconcile the expected and observed mass transfer requires substantial mass losses during the transfer between the donor star and the compact object (non-conservative mass transfer) as suggested for the AMXP SAX J1808.4$-$3658 \citep[see][for more details]{diSalvo08, Burderi09}. In line with this hypothesis, we note that the observed absorption column density ($N_H\simeq0.27\times 10^{22}$ cm$^{-2}$) is significantly larger than the averaged mapped value in the of direction of the globular cluster NGC 2808 \citep[$0.13 \times 10^{22}$ cm$^{-2}$;][]{Dickey1990a, Kalberla2005a}. This discrepancy, already reported for the AMXP SAX J1808.4$-$3658 \citep{Papitto09}, could in principle indicate the presence of additional neutral absorbers in the proximity of the source, in agreement with the ``radio-ejection'' scenario in which radiation pressure from the rotation-powered radio pulsar prevents Roche-Lobe overflow with consequent losses of matter \citep[see][for more details on the model]{Burderi02, diSalvo08}.

\begin{acknowledgements}
We thank the anonymous referee for helpful comments and suggestions that improved the paper.
We thank N. Schartel for the possibility to perform the ToO observation in the Director Discretionary Time, and the \xmm{} team for the technical support. We also use Director's Discretionary Time on \nustar{}, for which we thank Fiona Harrison for approving and the \nustar{} team for the technical support. We acknowledge financial contribution from the agreement ASI-INAF I/037/12/0. The High-Energy Astrophysics Group of Palermo acknowledges support from the Fondo Finalizzato alla Ricerca (FFR) 2012/13, project N. 2012- ATE-0390, founded by the University of Palermo. A. S. would like to thank F. Pintore for support and useful discussions. AP acknowledges support via an EU Marie Skodowska-Curie Individual fellowship under contract no. 660657-TMSP-H2020-MSCA-IF-2014, and the International Space Science Institute (ISSI) Bern, which funded and hosted the international team ``The disk-magnetosphere interaction around transitional millisecond pulsars''. NR acknowledges support from a Dutch NWO Vidi award A.2320.0076,  Spanish grants AYA2015-71042- P and SGR2014-1073, and the European COST Action MP1304 (NewCompstar).
\end{acknowledgements}

\bibliographystyle{aa} % style aa.bst
\bibliography{biblio_maxi}

\end{document}